\documentclass[10pt,conference,english,twocolumn]{IEEEtran}
\IEEEoverridecommandlockouts

\pdfoutput=1

\usepackage{babel}
\usepackage[babel]{microtype}
\usepackage[babel]{csquotes}

\usepackage[svgnames]{xcolor}
\usepackage{graphicx}

\usepackage{amsmath}
\usepackage{amsfonts}
\usepackage{amssymb}
\usepackage{amsthm}
\usepackage{algorithm,algpseudocode}
\usepackage{bm}
\usepackage{siunitx}
\usepackage{booktabs}
\usepackage{subfigure}
\usepackage{tabularx}
\usepackage{titlesec} 
\usepackage{geometry}
\usepackage{tikz}
\usepackage{pgfplots}
\pgfplotsset{compat=newest}
\usetikzlibrary{positioning}
\usetikzlibrary{shapes.geometric, arrows, decorations.pathreplacing}
\usetikzlibrary{shapes.arrows,arrows.meta}
\usetikzlibrary{3d}
\usetikzlibrary{calc}

\usepackage[defernumbers=false,
backend=biber,
isbn=false,
url=false,
maxbibnames=6,
maxcitenames=10,
doi=false,
giveninits=true,
sortcites=true,
eprint=false,
hyperref=true,
style=ieee,
bibstyle=ieee-proc,
]{biblatex}
\bibliography{ref.bib}

\usepackage[bookmarks=false]{hyperref}
\hypersetup{%
  pdftitle = {RISnet: A Domain-Knowledge Driven Neural Network Architecture for RIS Optimization with Mutual Coupling and Partial CSI},
  pdfauthor = {Bile Peng, Karl-Ludwig Besser, Shanpu Shen, Finn Siegismund-Poschmann, Ramprasad Raghunath, Daniel Mittleman, Vahid Jamali, Eduard A. Jorswieck},
  pdflang = {en-US},
  colorlinks,
  allcolors=.,
  urlcolor=MidnightBlue,
  bookmarksnumbered,
  bookmarksopen,
}

\usepackage[acronyms,nomain,xindy,nowarn]{glossaries}
\makeglossaries
\newacronym{isac}{ISAC}{integrated sensing and communication}
\newacronym{ae}{AE}{auto-encoder}
\newacronym{vae}{VAE}{variational auto-encoder}
\newacronym{ap}{AP}{access point}
\newacronym{aoa}{AoA}{angle-of-arrival}
\newacronym{ao}{AO}{alternating optimization}
\newacronym{bs}{BS}{base station}
\newacronym{dnn}{DNN}{deep neural network}
\newacronym{cnn}{CNN}{convolutional neural network}
\newacronym{rnn}{RNN}{recurrent neural network}
\newacronym{ma}{MA}{multiple access}
\newacronym{nn}{NN}{neural network}
\newacronym{ue}{UE}{user equipment}
\newacronym{noma}{NOMA}{non-orthogonal multiple access}
\newacronym{oma}{OMA}{orthogonal multiple access}
\newacronym{mimo}{MIMO}{multiple-input-multiple-output}
\newacronym{miso}{MISO}{multiple-input-single-output}
\newacronym{cf}{CF mMIMO}{cell-free massive MIMO}
\newacronym{cdf}{CDF}{cumulative distribution function}
\newacronym{gnn}{GNN}{graph neural network}
\newacronym{cc}{CC}{channel charting}
\newacronym{los}{LoS}{line-of-sight}
\newacronym{gmm}{GMM}{Gaussian mixture model}
\newacronym{csi}{CSI}{channel state information}
\newacronym{star}{STAR}{simultaneously transmitting and reflecting}
\newacronym{aod}{AoD}{angle-of-departure}
\newacronym{lwf}{LwF}{learing without forgetting}
\newacronym{sinr}{SINR}{signal-to-interference-plus-noise ratio}
\newacronym{snr}{SNR}{signal-to-noise ratio}
\newacronym{sgd}{SGD}{stochastic gradient descent}
\newacronym{agi}{AGI}{artificial general intelligence}
\newacronym{llm}{LLM}{large language model}
\newacronym{dl}{DL}{deep learning}
\newacronym{rl}{RL}{reinforcement learning}
\newacronym{crb}{CRB}{Cram\'er-Rao bound}
\newacronym{tpu}{TPU}{tensor processing unit}
\newacronym{evt}{EVT}{extreme value theory}
\newacronym{urllc}{URLLC}{ultra-reliable low latency communications}
\newacronym{mrt}{MRT}{maximum ratio transmission}
\newacronym{embb}{eMBB}{enhanced mobile broadband}
\newacronym{mmse}{MMSE}{minimum mean square error}
\newacronym{qd}{QD}{quasi degradation}
\newacronym{rss}{RSS}{received signal strength}
\newacronym{siso}{SISO}{single-input single-output}
\newacronym{rsma}{RSMA}{rate-splitting multiple access}
\newacronym{lstm}{LSTM}{long short-term memory}
\newacronym{drl}{DRL}{deep reinforcement learning}
\newacronym{dqn}{DQN}{deep Q-network}
\newacronym{ml}{ML}{machine learning}
\newacronym{sca}{SCA}{successive convex approximation}
\newacronym{admm}{ADMM}{alternating direction method of multipliers}
\newacronym{mm}{MM}{majorization-minimization}
\newacronym{rmcg}{RMCG}{Riemannian manifold conjugate gradient}
\newacronym{sdma}{SDMA}{space-division multiple access}
\newacronym{sdr}{SDR}{semidefinite relaxation}
\newacronym{mse}{MSE}{mean squared error}
\newacronym{ris}{RIS}{reconfigurable intelligent surface}
\newacronym{tof}{ToF}{time of flight}
\newacronym{em}{EM}{expectation-maximization}
\newacronym{zf}{ZF}{zero-forcing}
\newacronym{ekf}{EKF}{extended Kalman filter}
\newacronym{bcd}{BCD}{block coordinate descent}
\newacronym{concrete}{CONCRETE}{CONtinuous relaxation for disCRETE}
\newacronym{mop}{MOP}{multi-objective programming}
\newacronym{uwb}{UWB}{ultra-wide band}
\newacronym{sic}{SIC}{successive interference cancellation}
\newacronym{otfs}{OTFS}{orthogonal time frequency space}
\newacronym{ofdm}{OFDM}{orthogonal frequency-division multiplexing}
\newacronym{mpc}{MPC}{multi-path component}
\newacronym{wmmse}{WMMSE}{weighted minimum mean square error}
\newacronym{wsr}{WSR}{weighted sum-rate}
\newacronym{mab}{MAB}{multi-armed bandit}
\newacronym{ppo}{PPO}{proximal policy optimization}
\newacronym{sac}{SAC}{soft actor-critic}
\newacronym{iid}{i.i.d.}{independent and identically distributed}
\newacronym{wrt}{w.r.t.}{with respect to}
\newacronym{fdma}{FDMA}{frequency domain multiple access}
\newacronym{tdma}{TDMA}{time domain multiple access}
\newacronym{cdma}{CDMA}{code domain multiple access}
\newacronym{swipt}{SWIPT}{simultaneous wireless information and power transfer}
\setacronymstyle{long-short}
\glsdisablehyper

\addto\extrasenglish{%

}
\addto\extrasenglish{%

}

\theoremstyle{remark}
\newtheorem{remark}{Remark}

\theoremstyle{definition}

\IfPackageLoadedTF{titlesec}{%
	\ifCLASSOPTIONconference%
	\titlespacing{\section}{0pt}{1.5ex plus 1.5ex minus 0.5ex}{0.7ex plus 1ex minus 0ex} 
	\titlespacing{\subsection}{0pt}{1.5ex plus 1.5ex minus 0.5ex}{0.7ex plus .5ex minus 0ex} 
	\else
	\titlespacing{\section}{0pt}{3.0ex plus 1.5ex minus 1.5ex}{0.7ex plus 1ex minus 0ex} 
	\titlespacing{\subsection}{0pt}{3.5ex plus 1.5ex minus 1.5ex}{0.7ex plus .5ex minus 0ex} 
	\fi
	
	\def\thesubsubsectiondis{\arabic{subsubsection})}
	\def\theparagraphdis{\alph{paragraph})}
	\titleformat{\subsubsection}[runin]{\normalfont\normalsize\itshape}{\thesubsubsectiondis}{.5em}{}[:]
	\titlespacing*{\subsubsection}{\parindent}{0ex plus 0.1ex minus 0.1ex}{1ex}
	\titleformat{\paragraph}[runin]{\normalfont\normalsize\itshape}{\theparagraphdis}{.5em}{}[:]
	\titlespacing*{\paragraph}{2\parindent}{0ex plus 0.1ex minus 0.1ex}{1ex}
}{}

\newcommand{\bPhi}{\boldsymbol{\Phi}}

\newcommand{\e}{\mathrm{e}}

\let\imag\jj

\newcommand{\norm}[1]{\lVert#1\rVert}

\newcommand{\todo}[2][]{\ignorespaces
	\if\relax\detokenize{#1}\relax
	{\color{red}[TODO: #2]}%
	\else
	{\color{red}[TODO (#1): #2]}%
	\fi
}

\definecolor{plot0}{HTML}{004488}
\definecolor{plot1}{HTML}{DDAA33}
\definecolor{plot2}{HTML}{BB5566}
\definecolor{plot3}{HTML}{000000}
\definecolor{plot4}{HTML}{AAAAAA}

\title{RIS-Assisted NOMA with Partial CSI and Mutual Coupling: A Machine Learning Approach}
\author{
\IEEEauthorblockN{%
Bile Peng\IEEEauthorrefmark{1}, 
Karl-Ludwig Besser\:\!\IEEEauthorrefmark{2}, 
Shanpu Shen\IEEEauthorrefmark{3}, 
Finn Siegismund-Poschmann\IEEEauthorrefmark{4},
Ramprasad Raghunath\IEEEauthorrefmark{1},\\
Daniel M. Mittleman\IEEEauthorrefmark{5},
Vahid Jamali\IEEEauthorrefmark{6},
and
Eduard A. Jorswieck\IEEEauthorrefmark{1} 
}

 \IEEEauthorblockA{\IEEEauthorrefmark{1}Institute for Communications Technology, Technische Universit\"at Braunschweig, Germany}

 \IEEEauthorblockA{\IEEEauthorrefmark{2}Department of Electrical Engineering, Linköping University, Linköping, Sweden}

 \IEEEauthorblockA{\IEEEauthorrefmark{3}State Key Laboratory of Internet of Things for Smart City, University of Macau, Macau, China}

 \IEEEauthorblockA{\IEEEauthorrefmark{4}Institute for Computer Science, Freie Universität Berlin, Germany}

 \IEEEauthorblockA{\IEEEauthorrefmark{5}School of Engineering, Brown University, USA}

 \IEEEauthorblockA{\IEEEauthorrefmark{6}Department of Electrical Engineering and Information Technology, TU Darmstadt, Germany}

\thanks{The work of E.~Jorswieck, R.~Raghunath, and B.~Peng was supported partly by the Federal Ministry of Education and Research (BMBF), Germany, through the Program of Souverän, Digital, and Vernetzt Joint Project 6G-RIC under Grant~16KISK031.
The work of B.~Peng is partly supported by the German research foundation (DFG) as part of the ML4RIS project (566937681).
The work of E.~Jorswieck received partly support from the Smart Networks and Services Joint Undertaking (SNS JU) under the
European Union’s Horizon Europe research and innovation programme within 6G-SENSES project (Grant Agreement No 101139282).
The work of K.-L.~Besser is supported by the Security Link project.
The work of S. Shen is funded by The Science and Technology Development Fund, Macau SAR (File/Project no. 001/2024/SKL and CG2025/IOTSC)
The work of V.~Jamali is supported in part by the DFG within the Collaborative Research Center MAKI (SFB~1053, Project-ID~210487104) and in part by the LOEWE initiative (Hesse, Germany) within the emergenCITY center [LOEWE/1/12/519/03/05.001(0016)/72].
The work of D.~Mittleman is supported by the US National Science Foundation (CNS-1954780, CNS-2211616), the US Air Force Office of Scientific Research (FA9550-22–1-0412), the Alexander von Humboldt Foundation, 
and the DFG through a Mercator Fellowship.}
}

\begin{document}

\maketitle

\begin{abstract}
\Gls{noma} is a promising multiple access technique.
Its performance depends strongly on the wireless channel property,
which can be enhanced by \glspl{ris}.
In this paper,
we jointly optimize \gls{bs} precoding and \gls{ris} configuration with
unsupervised \gls{ml},
which looks for the optimal solution autonomously.
In particular, we propose a dedicated \gls{nn} architecture RISnet
inspired by domain knowledge in communication.
Compared to state-of-the-art,
the proposed approach combines analytical optimal \gls{bs} precoding and \gls{ml}-enabled \gls{ris},
has a high scalability to control more than 1000 \gls{ris} elements,
has a low requirement for \gls{csi} in input,
and addresses the mutual coupling between \gls{ris} elements.
Beyond the considered problem,
this work is an early contribution to domain knowledge enabled \gls{ml},
which exploit the domain expertise of communication systems to design better approaches than general \gls{ml} methods.
\end{abstract}


\glsresetall
\section{Introduction}
\label{sec:intro}

\Gls{noma} is a promising technique to serve multiple users simultaneously.
Compared to \gls{oma}, 
\gls{noma} realizes a higher spectrum efficiency because it uses the same radio resource to serve multiple users.
Compared to \gls{sdma},
\gls{noma} realizes better fairness,
and does not require a high rank channel matrix~\cite{liu2022evolution}.

The performance of \gls{noma} relies on the channel property,
for example, the channel degradation~\cite{jorswieck2021optimality}
or quasi-degradation~\cite{Chen22016}.
In recent years,
the \gls{ris} has drawn significant attention from both academia and industry 
due to its ability to manipulate the channel property.
It comprises many passive antenna elements.
Each element receives signals from the transmitter
(e.g., the \gls{bs}),
performs a simple signal processing without external power
(e.g., phase shift),
and transmits it to the receivers
(e.g., the users).
The large number of \gls{ris} elements allows for a high flexibility,
which makes it suitable to realize favorable channel properties for \gls{noma}~\cite{gao2021machine,guo2022energy}.

In the literature,
\gls{noma} precoding and \gls{ris} configuration are
jointly optimized with \gls{sdr}~\cite{fu2019intelligent,yang2021reconfigurable}.
\Gls{ris} partitioning is proposed for \gls{noma}~\cite{khaleel2021novel}.
A common limitation of the above listed references
is the scalability:
Most of them assume
no more than 100 \gls{ris} elements,
far from the vision of thousands of \gls{ris} elements~\cite{liu2021reconfigurable}
and the requirement to achieve a reasonable link budget.
Another common limitation of them is the assumption of known \gls{csi}.
Due to the large number of \gls{ris} elements,
the full \gls{csi} is very expensive to acquire.
Moreover, most existing references assume a perfect \gls{ris} without mutual coupling.
However,
due to the small distance between two adjacent \gls{ris} elements,
there might exist mutual coupling between them.
Although \gls{ris} with mutual coupling is modeled and optimized in the literature~\cite{pettanice2023mutual,shen2021modeling,Peng2025risnet},
the joint optimization of the \gls{noma} precoding and \gls{ris} configuration
considering mutual coupling
is still an open problem.

In this paper, we jointly optimize the \gls{noma} precoding and \gls{ris} configuration
with unsupervised \gls{ml},
where the \gls{nn} looks for the optimal solution autonomously without given labels.
Unlike our previous work on \gls{ris} for \gls{sdma},
we propose a dedicated \gls{nn} architecture \emph{RISnet} for \gls{noma} with a high scalability to control more than
1000 \gls{ris} elements,
a low requirement for \gls{csi},
and the ability to consider mutual coupling in \gls{ris} explicitly.

\section{Problem Formulation}
\label{sec:problem}

\begin{figure}[htbp]
    \centering
    \resizebox{.5\linewidth}{!}{
    \begin{tikzpicture}
    \tikzstyle{base}=[isosceles triangle, draw, rotate=90, fill=gray!60, minimum size =.5cm]
	\tikzstyle{user}=[rectangle, draw, fill=gray!60, minimum size =.5cm, rounded corners=0.1cm]
	\tikzstyle{element}=[rectangle, fill=gray!30]
	
	\node[base,label={left:BS}] (BS) at (-3,0){};
	\draw[decoration={expanding waves,segment length=6},decorate] (BS) -- (-3,1.5);
	\node[user,label={below:User 1}] (UE1) at (4,2){};
	\node[user,label={below:User 2}] (UE2) at (4,0){};
	\draw[step=0.33cm,thick] (-1,1.98) grid (0, 3);
	\node[label={[label distance=10]above:RIS}] (RIS) at (-0.5, 2.5) {};
	
	\node[above right=.15 and .3 of RIS]{$\boldsymbol{\Phi}$};
	
	\draw[-to,shorten >=3pt] (BS) to node[above left, pos=.1, yshift=-.7cm] {$\mathbf{V}$} (UE1);
	\draw[-to,shorten >=3pt] (BS) to node[above,pos=.3, ] {$\mathbf{D}$} (UE2);
	
	\draw[-to] (BS) to node[above left, pos=.6] {$\mathbf{H}$} (RIS);
	
	\draw[-to,shorten >=3pt] (RIS) to node[left=-1mm, below=0mm, pos=.3] {$\mathbf{G}$} (UE1);
	\draw[-to,shorten >=3pt] (RIS) to node[below=-3mm, left=2mm] {} (UE2);
\end{tikzpicture}}
    \caption{The system model of \gls{ris}-assisted downlink multi-user broadcasting.}
    \label{fig:system_model}
\end{figure}
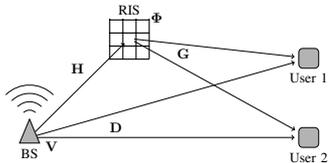

We consider a \gls{ris}-aided two-user \gls{miso} scenario,
as illustrated in \autoref{fig:system_model}.
The channel from \gls{bs} to \gls{ris} is $\mathbf{H} \in \mathbb{C}^{N\times M}$,
where $N$ is the number of \gls{ris} elements
and $M$ is the number of \gls{bs} antennas.
The channel from \gls{ris} to users is $\mathbf{G} \in \mathbb{C}^{2\times N}$.
The two rows of $\mathbf{G}$ corresponds to the \gls{miso} channels to two users.
The direct channel from \gls{bs} to users directly is $\mathbf{D} \in \mathbb{C}^{2\times M}$,
where the two rows correspond to channels to two users as well.

\begin{remark}
We assume two users in this scenario following typical problem formulations in \gls{noma}~\cite{Chen22016,zhu2020power,jorswieck2021optimality}.
This assumption can be generalized to more than two users with, e.g., user pairing.
\end{remark}

Denote the precoding matrix as $\mathbf{V} \in \mathbb{C}^{M\times 2}$
and the signal processing matrix of the \gls{ris} as the diagonal matrix~${\boldsymbol{\Phi} \in \mathbb{C}^{N\times N}}$, 
where the diagonal element in row~$n$ and column~$n$ is $\phi_{nn}=\e^{\imag\psi_n}$, 
with $\psi_n$ being the phase shift of \gls{ris} element~$n$. 
Taking the mutual coupling in the \gls{ris} into consideration,
the channel~$\mathbf{C}$ between \gls{bs} and users is~\cite{Peng2025risnet}
\begin{equation}
    \mathbf{C} = \mathbf{D} + \mathbf{G} (\mathbf{I} - \boldsymbol{\Phi}\mathbf{S}_{II})^{-1}\boldsymbol{\Phi}\mathbf{H},
    \label{eq:channel_mutual_coupling}
\end{equation}
where $\mathbf{S}_{II}$ is the S-parameter matrix of the \gls{ris}. 
If the \gls{ris} does not have mutual coupling,
$\mathbf{S}_{II}=\mathbf{0}$ and
\eqref{eq:channel_mutual_coupling} is reduced to
$\mathbf{C} = \mathbf{D} + \mathbf{G} \boldsymbol{\Phi}\mathbf{H}$~\cite{guo2020weighted}.
The signal received at the users is
\begin{equation}
\mathbf{y} =  \mathbf{C}\mathbf{V} \mathbf{x} + \mathbf{n},
\label{eq:transmission_los}
\end{equation}
where $\mathbf{x} \in \mathbb{C}^{2 \times 1}$ is the symbols to be transmitted,
$\mathbf{y} \in \mathbb{C}^{2 \times 1}$ is the received symbols, and $\mathbf{n} \in \mathbb{C}^{2 \times 1}$ is the noise.

Without loss of generality,
we assume user~1 has a higher channel gain than user~2.
Following the canonical problem formulation of \gls{noma}~\cite{chen2016optimal},
we aim to minimize the transmit power subject to the rate requirements,
 $r_1$ and $r_2$ of user~1 and user~2, respectively:
\begin{subequations}
\begin{alignat}{2}
&\!\min_{\mathbf{v}_1, \mathbf{v}_2, \bPhi} &\quad& f = \norm{\mathbf{v}_1}^2 + \norm{\mathbf{v}_2}^2,   \label{eq:optProbMinPW}\\
&\text{s.t.} &      & \beta_1 \geq 2^{r_1} - 1,  \label{eq:optProbMinPWconstraint1}\\
&                  &      & \min \left\{\beta_{21}, \beta_{22} \right\} \geq 2^{r_2} - 1,   \label{eq:optProbMinPWconstraint2}
\end{alignat}%
\label{eq:problem_noma}%
\end{subequations}
where $\mathbf{v}_1$ and $\mathbf{v}_2$ are precoding vectors for the two users, $\beta_{21}$ is the \gls{sinr} of the signal for user~2 at user~1, 
$\beta_1$ is the \gls{snr} of the signal for user~1 at user~1,
and $\beta_{22}$ is the \gls{sinr} of the signal for user~2 at user~2.

\section{Unsupervised Machine Learning with RISnet}
\label{sec:risnet}

\subsection{The Framework of Unsupervised ML for Optimization}
\label{sec:framework}

In this section,
we first present the framework of unsupervised \gls{ml} for optimization.
Given a problem representation~$\boldsymbol{\Gamma}$ (in our case, \gls{csi} and required rates~$r_1$ and $r_2$ in~\eqref{eq:problem_noma}),
we look for a solution~$\boldsymbol{\Phi}$ (the \gls{ris} phase shifts) that minimizes objective~$f$ in~\eqref{eq:optProbMinPW},
which is fully determined by $\boldsymbol{\Gamma}$ and $\boldsymbol{\Phi}$, and it can be written as
$f(\boldsymbol{\Gamma}, \boldsymbol{\Phi})$.
We define an \gls{nn}~$N_\theta$, which is parameterized by~$\theta$ (i.e., $\theta$ contains all trainable weights and biases in~$N_\theta$) and
maps from~$\boldsymbol{\Gamma}$
to~$\boldsymbol{\Phi}$,
i.e., $\boldsymbol{\Phi} = N_\theta(\boldsymbol{\Gamma})$.
We write the objective as ${f(\boldsymbol{\Gamma}, \boldsymbol{\Phi})=f(\boldsymbol{\Gamma}, N_\theta(\boldsymbol{\Gamma}); \theta)}$.
Note that it is emphasized that~$f$ depends on~$\theta$.
We then collect massive data of $\boldsymbol{\Gamma}$ in a training set~$\mathcal{D}$
and formulate the problem as
\begin{equation}
    \min_\theta K=\sum_{\boldsymbol{\Gamma} \in \mathcal{D}} f(\boldsymbol{\Gamma}, N_\theta(\boldsymbol{\Gamma}); \theta).
    \label{eq:ml}
\end{equation}
In this way, $N_\theta$ is optimized for the emsemble of $\boldsymbol{\Gamma} \in \mathcal{D}$
(\emph{training}) using gradient descent:
\begin{equation}
    \theta \leftarrow \theta - \eta \nabla_\theta K,
    \label{eq:gradient-ascend}
\end{equation}
where $\eta$ is the learning rate.
If $N_\theta$ is well trained,
$\boldsymbol{\Phi}' = N_\theta (\mathbf{\Gamma'})$ is also a good solution for $\boldsymbol{\Gamma}' \notin \mathcal{D}$ (\emph{testing}),
like a human uses experience to solve new problems of the same type\footnote{A complete retraining is only required when the input states are fundamentally changed, e.g., change of deployment environment.}~\cite{yu2022role}.

If $\boldsymbol{\Gamma}$ (e.g., the full \gls{csi}) is difficult to obtain,
we use an observation $\mathbf{O}$ of $\boldsymbol{\Gamma}$
(e.g., the partial \gls{csi})
as the input of $N_\theta$,
i.e., $\boldsymbol{\Phi} = N_\theta(\mathbf{O})$.
It facilitates the application of $N_\theta$ once it is trained.
However, $\boldsymbol{\Gamma}$ might not be fully determined given $\mathbf{O}$.
Therefore, the optimization might be more difficult.

Although~\eqref{eq:ml} is a general approach,
it would benefit from the problem-specific domain knowledge.
In the following sections,
we first define the \gls{nn} input.
Next, we propose the RISnet architecture.
\subsection{The RISnet Architecture with Full CSI}
\label{sec:risnet_architecture}

The RISnet is designed according to our understanding of the problem property.
The phase shift of a \gls{ris} element
depends on both its own channel gain
and the common optimization objective of the \gls{ris} array.
Correspondingly,
we define a \emph{local feature} for each \gls{ris} element,
and a \emph{global feature} shared by all \gls{ris} elements.
In a RISnet of $I$ layers,
the output of layer~$i$ ($i>1$)
is computed using both local and global features in the input,
such that the above-mentioned problem property is reflected in the inference of the \gls{nn}.
In layer~1,
we define input feature per \gls{ris} element.
It is straightforward to include channel gains from \gls{ris} element~$n$ to users
(i.e., column~$n$ of $\mathbf{G}$)
as well as the rate requirements $r_1$ and $r_2$
in the input feature.
However,
the direct channel~$\mathbf{D}$ does not involve the \gls{ris} and its columns cannot be mapped to 
\gls{ris} elements.
As a workaround, we define
$\mathbf{J}=\mathbf{D}\mathbf{H}^+$,
where $\mathbf{H}^+$ is the pseudo-inverse of $\mathbf{H}$.
Then, the signal transmission is
\begin{equation}
\mathbf{y} = \left(\mathbf{G} (\mathbf{I}-\boldsymbol{\Phi}\mathbf{S}_{II})^{-1} \boldsymbol{\Phi} + \mathbf{J}\right) \mathbf{H} \mathbf{V} \mathbf{x} + \mathbf{n}.
\label{eq:transformed_channel}
\end{equation}
We can inteprete \eqref{eq:transformed_channel} as follows:
signal~$\mathbf{x}$ is precoded with~$\mathbf{V}$, transmitted through channel~$\mathbf{H}$ to the \gls{ris},
and through channel~${\mathbf{G}(\mathbf{I}-\boldsymbol{\Phi}\mathbf{S}_{II})^{-1}\boldsymbol{\Phi} + \mathbf{J}}$
to the users. 
Element~$j_{un}$ of~$\mathbf{J}$ can be interpreted as the channel gain from \gls{ris} element~$n$ to user~$u$.
Therefore,
the feature of \gls{ris} element~$n$ is defined as
\begin{equation}
\begin{aligned}
\boldsymbol{\gamma}_{n}
=  ( &
|g_{1n}|, \arg(g_{1n}), |j_{1n}|, \arg(j_{1n}), r_1, \\
& |g_{2n}|, \arg(g_{2n}), |j_{2n}|, \arg(j_{2n}), r_2
)^T \in \mathbb{R}^{10 \times 1}.
\end{aligned}
\end{equation}
The entire input feature of all $N$ \gls{ris} elements is the concatenation of features per \gls{ris} element.
Therefore, it has the shape of $10 \times N$.

Denote the input feature of \gls{ris} element~$n$ in layer~$i$ as $\mathbf{f}_{n,i}$
and define classes of \texttt{c}urrent \gls{ris} element~\enquote{\texttt{c}} (for \emph{current})
and \texttt{a}ll elements~\enquote{\texttt{a}} (for \emph{all}).
Applying the idea of local and global features,
the output feature of \gls{ris} element~$n$ in layer~$i$ is
\begin{equation}
\mathbf{f}_{n, i + 1} =
\begin{pmatrix}
     \text{ReLU}(\mathbf{W}^{\texttt{c}}_{i} \mathbf{f}_{n, i} + \mathbf{b}_i^{\texttt{c}}) \\
     \big(\sum_{n'}\text{ReLU}(\mathbf{W}^{\texttt{a}}_{i} \mathbf{f}_{n', i} + \mathbf{b}_i^{\texttt{a}})\big) \big/ N\\
\end{pmatrix}
\label{eq:layer_processing_noma}
\end{equation}
for $i<L$, where $\mathbf{W}^c_i \in \mathbb{R}^{Q_i\times P_i}$ are the trainable weights of class~c in layer~$i$ with the input feature dimension~$P_i$ in layer~$i$ (i.e., $\mathbf{f}_{n, i} \in \mathbb{R}^{P_i \times 1}$) and output feature dimension~$Q_i$ in layer~$i$ of class~\texttt{c},
$\mathbf{b}^{c}_i \in \mathbb{R}^{Q_i \times 1}$ is trainable bias of class~\texttt{c} in layer~$i$.
Similar definitions and same dimensions apply to class~\texttt{a}.
It is also straightforward to infer $P_{i+1}=2Q_i$.
Observe~\eqref{eq:layer_processing_noma},
we note that the same information processing units are applied to all users and \gls{ris} elements.
Therefore,
the number of trainable parameters is independent from the number of \gls{ris} elements,
which enables a high scalability to configure more than 1000 \gls{ris} elements.
For the final layer,
we use one information processing unit, i.e.,
\begin{equation}
\varphi_{n} =
     \text{ReLU}(\mathbf{W}^{\texttt{c}}_{I} \mathbf{f}_{n, I} + \mathbf{b}_I^{\texttt{c}}).
\label{eq:layer_processing_noma_final}
\end{equation}

Element~$\phi_{nn}$ in row~$n$ and column~$n$ of $\boldsymbol{\Phi}$ is defined as
$\phi_{nn}=\e^{\imag\varphi_n}$.

\begin{figure}
    \centering
    \subfigure[First layer]{\resizebox{.9\linewidth}{!}{\begin{tikzpicture}
\tikzstyle{layer} = [rectangle, rounded corners, minimum width=2.5cm, minimum height=.6cm, align=center, text centered, draw=black]

\draw[thick] (0, 0) grid +(4, 2);

\node[font=\LARGE] at (2, -.5) {RIS Element};
\node[font=\LARGE] at (-2.25, 1.5) {Feature of user~1};
\node[font=\LARGE] at (-2.25, .5) {Feature of user~2};

\draw [decorate,decoration={brace,amplitude=5pt,mirror}]
(-4.25,-1) -- (4.25,-1) node[midway, yshift=-.7cm, font=\LARGE]{Input tensor};
\draw [decorate,decoration={brace,amplitude=5pt,mirror}]
(4.75,-1) -- (7.5,-1) node[midway, yshift=-.7cm, font=\LARGE]{Info. proc.};
\draw [decorate,decoration={brace,amplitude=5pt,mirror}]
(8,-1) -- (15.1,-1) node[midway, yshift=-.7cm, font=\LARGE]{Output tensor};

\node[draw,thick,rounded corners,rotate=90,inner sep=0em,minimum width=2cm,minimum height=1.5cm] (filters) at (6., 1) {};
\node (layer1) [xshift=6cm, yshift=1.5cm,font=\LARGE] {\texttt{c}};
\node (layer2) [below of=layer1, yshift=0cm,font=\LARGE] {\texttt{a}};
\draw [->,thick] (layer1.east) -- (7.75, 1.5);
\draw [->,thick] (layer2.east) -- (7.75, 0.5);
\draw [->,thick] (4.25, 1) -- ++ (.95, 0);

\draw[thick] (11, 0) grid +(4, 2);
\node[font=\LARGE] at (13, -.5) {RIS Element};
\node[font=\LARGE] at (9.5, 1.5) {Feature \texttt{c}};
\node[font=\LARGE] at (9.5, .5) {Feature \texttt{a}};
\end{tikzpicture}}}
    \subfigure[Intermediate layers]{\resizebox{.9\linewidth}{!}{\begin{tikzpicture}
\tikzstyle{layer} = [rectangle, rounded corners, minimum width=2.5cm, minimum height=.6cm, align=center, text centered, draw=black]

\draw[thick] (0, 0) grid +(4, 2);

\node[font=\LARGE] at (2, -.5) {RIS Element};
\node[font=\LARGE] at (-1.5, 1.5) {Feature \texttt{c}};
\node[font=\LARGE] at (-1.5, .5) {Feature \texttt{a}};

\draw [decorate,decoration={brace,amplitude=5pt,mirror}]
(-3,-1) -- (4.25,-1) node[midway, yshift=-.7cm, font=\LARGE]{Input tensor};
\draw [decorate,decoration={brace,amplitude=5pt,mirror}]
(4.75,-1) -- (7.5,-1) node[midway, yshift=-.7cm, font=\LARGE]{Info. proc.};
\draw [decorate,decoration={brace,amplitude=5pt,mirror}]
(8,-1) -- (15.1,-1) node[midway, yshift=-.7cm, font=\LARGE]{Output tensor};

\node[draw,thick,rounded corners,rotate=90,inner sep=0em,minimum width=2cm,minimum height=1.5cm] (filters) at (6., 1) {};
\node (layer1) [xshift=6cm, yshift=1.5cm,font=\LARGE] {\texttt{c}};
\node (layer2) [below of=layer1, yshift=0cm,font=\LARGE] {\texttt{a}};
\draw [->,thick] (layer1.east) -- (7.75, 1.5);
\draw [->,thick] (layer2.east) -- (7.75, 0.5);
\draw [->,thick] (4.25, 1) -- ++ (.95, 0);

\draw[thick] (11, 0) grid +(4, 2);
\node[font=\LARGE] at (13, -.5) {RIS Element};
\node[font=\LARGE] at (9.5, 1.5) {Feature \texttt{c}};
\node[font=\LARGE] at (9.5, .5) {Feature \texttt{a}};
\end{tikzpicture}}}
    \subfigure[Final layer]{\resizebox{.9\linewidth}{!}{\begin{tikzpicture}
\tikzstyle{layer} = [rectangle, rounded corners, minimum width=2.5cm, minimum height=.6cm, align=center, text centered, draw=black]

\draw[thick] (0, 0) grid +(4, 2);

\node[font=\LARGE] at (2, -.5) {RIS Element};
\node[font=\LARGE] at (-1.5, 1.5) {Feature \texttt{c}};
\node[font=\LARGE] at (-1.5, .5) {Feature \texttt{a}};

\draw [decorate,decoration={brace,amplitude=5pt,mirror}]
(-2.75,-1) -- (4.5,-1) node[midway, yshift=-.7cm, font=\LARGE]{Input tensor};
\draw [decorate,decoration={brace,amplitude=5pt,mirror}]
(4.75,-1) -- (9.25,-1) node[midway, yshift=-.7cm, font=\LARGE]{Info. proc.};
\draw [decorate,decoration={brace,amplitude=5pt,mirror}]
(9.5,-1) -- (16.1,-1) node[midway, yshift=-.7cm, font=\LARGE]{Output tensor};

\node (layer1) [draw,rectangle,rounded corners,thick,inner sep=.25cm,font=\LARGE, align=left,anchor=west] at (5.5, 1) {Information\\Processing};
\draw [->,thick] (layer1.east) -- (10, 1);
\draw [->,thick] (4.5, 1) -- (layer1.west);


\begin{scope}[yshift=-.5cm]
    

\draw (12, 1) grid +(4, 1);
\node[font=\LARGE] at (14, 0) {RIS element};
\node[font=\LARGE, text width=2cm] at (11.2, 1.5) {Phase shifts};
\end{scope}

\end{tikzpicture}}}
    \caption{Information processing of RISnet.
    }
    \label{fig:info_processing_noma}
\end{figure}
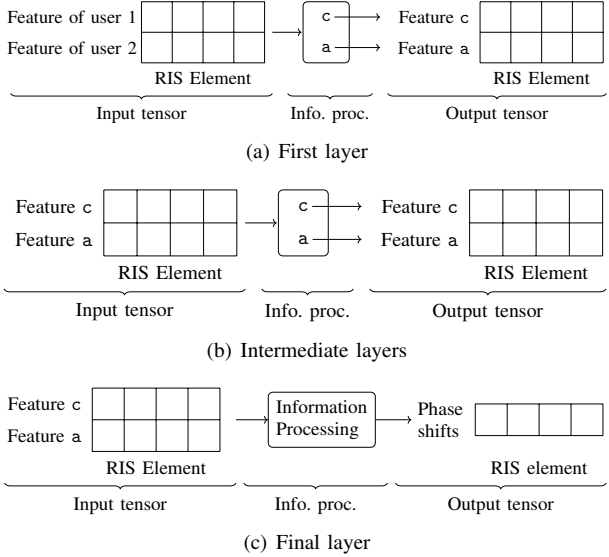

\subsection{The RISnet Architecture with Partial CSI}

With the above-described RISnet architecture,
we realize a high scalability.
However, it still requires the full \gls{csi} as input,
which is very expensive to acquire due to the many \gls{ris} elements.
As a countermeasure, we propose to use partial \gls{csi} of only a few \gls{ris} elements as the input of RISnet
instead of the full \gls{csi}.
This partial \gls{csi} is defined as the observation~$\mathbf{O}$ of the full \gls{csi}~$\boldsymbol{\Gamma}$.
If the channel has spatial correlation~\cite{pizzo2020spatially},
i.e., there are a few specular propagation paths
rather than infinitely many infinitely weak paths,
we can infer $\boldsymbol{\Gamma}$ from $\mathbf{O}$ with some uncertainty.
Instead of inferring $\boldsymbol{\Gamma}$ explicitly,
we learn an end-to-end mapping from partial \gls{csi} directly to the \gls{ris} configuration
in order to reduce the algorithm complexity.

\begin{figure}
    \centering
    \tikzset{
	anchors1/.pic={
		\draw[step=1, lightgray] (0,0) grid (36, 36);
		\foreach \k in {0,...,3}
		\foreach \l in {0,...,3}
		{
			\fill[plot0] (9*\k+4, 9*\l+4) rectangle (9*\k+5, 9*\l+5);
		}
	},
	anchors2/.pic={
		\draw[step=1, lightgray] (0,0) grid (36, 36);
		\draw[step=9, gray] (0,0) grid (36, 36);
		\foreach \k in {0,...,3}
		\foreach \l in {0,...,3}
		{
		\foreach \i in {0,...,2}
		\foreach \j in {0,...,2}
		{
				\fill[plot1] (3*\i+1 + 9*\k, 3*\j+1 + 9*\l) rectangle (3*\i+2 + 9*\k, 3*\j + 2 + 9*\l);
		}
		\fill[plot0] (9*\k+4, 9*\l+4) rectangle (9*\k+5, 9*\l+5);
		}
	},
	anchors3/.pic={
		\draw[fill=plot1,fill opacity=.5] (0, 0) rectangle (36, 36);
		\draw[step=1, lightgray] (0,0) grid (36, 36);
		\draw[step=3, gray] (0,0) grid (36, 36);
		\foreach \k in {0,...,3}
		\foreach \l in {0,...,3}
		\foreach \i in {0,...,2}
		\foreach \j in {0,...,2}
		{
			\fill[plot0] (3*\i+1 + 9*\k, 3*\j+1 + 9*\l) rectangle (3*\i+2 + 9*\k, 3*\j + 2 + 9*\l);
		}
	}
}
\begin{tikzpicture}
\tikzstyle{layer} = [rectangle, rounded corners, minimum width=2.5cm, minimum height=.6cm, align=center, draw=black]

\node (input) [layer] {Observation $\mathbf{O}$};
\node (layer1) [layer,below=.5 of input] {Normal layer (16 RIS elements)};
\node (layer2) [layer, below=.1 of layer1] {Normal layer (16 RIS elements)};
\node (layer3) [layer, below=1 of layer2] {Expansion layer (144 RIS elements)};
\node (layer4) [layer, below=.1 of layer3] {Normal layer (144 RIS elements)};
\node (layer5) [layer, below=.1 of layer4] {Normal layer (144 RIS elements)};
\node (layer6) [layer, below=1 of layer5] {Expansion layer (1296 RIS elements)};
\node (layer7) [layer, below=.1 of layer6] {Normal layer (1296 RIS elements)};
\node (layer8) [layer, below=.1 of layer7] {Normal layer (1296 RIS elements)};
\node (output) [layer, below=.5 of layer8] {Solution $\boldsymbol{\Phi}$};

\pic[scale=.08] at ($(layer2)!.5!(layer3) + (3, 0)$) {anchors1};
\pic[scale=.08] at ($(layer5)!.5!(layer6) + (3, 0)$) {anchors2};
\pic[scale=.08] at ($(output.north) + (3, 0)$) {anchors3};

\draw[->] (input) -- (layer1);
\draw[->] (layer8) -- (output);
\draw[->] (layer2) -- (layer3);
\draw[->] (layer5) -- (layer6);
\end{tikzpicture}
    \caption{Expansion of considered \gls{ris} elements using two expansion layers.
    In the RIS illustrations (right),
    blue elements have been taken into consideration in RISnet,
    yellow elements are to be considered in the current expansion layer,
    white elements are not yet considered.
    }
    \label{fig:risnet_partial}
\end{figure}
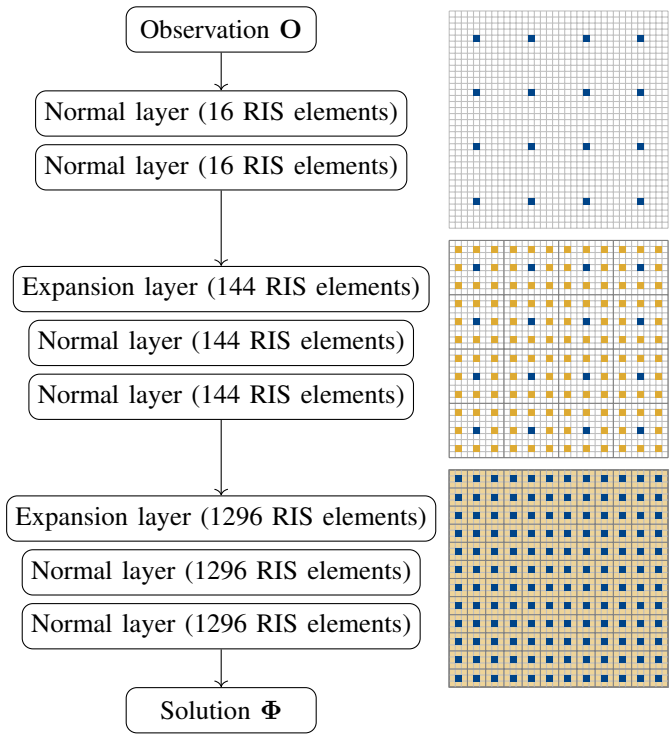

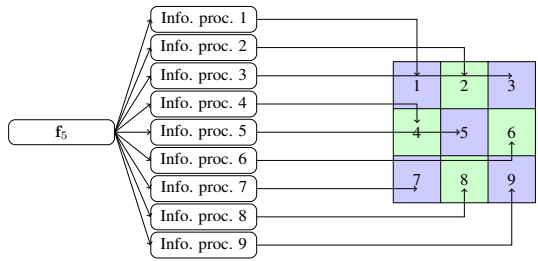
\begin{figure}
    \centering
    \resizebox{.8\linewidth}{!}{
        \begin{tikzpicture}
\tikzstyle{block} = [rectangle, rounded corners, text width=2cm, text centered, draw=black]
    
\node (feature) [block] {$\mathbf{f}_5$};
\node (filter1) [block, right of=feature, xshift=2cm, yshift=2.4cm] {Info. proc. 1};
\node (filter2) [block, below of=filter1, yshift=.4cm] {Info. proc. 2};
\node (filter3) [block, below of=filter2, yshift=.4cm] {Info. proc. 3};
\node (filter4) [block, below of=filter3, yshift=.4cm] {Info. proc. 4};
\node (filter5) [block, below of=filter4, yshift=.4cm] {Info. proc. 5};
\node (filter6) [block, below of=filter5, yshift=.4cm] {Info. proc. 6};
\node (filter7) [block, below of=filter6, yshift=.4cm] {Info. proc. 7};
\node (filter8) [block, below of=filter7, yshift=.4cm] {Info. proc. 8};
\node (filter9) [block, below of=filter8, yshift=.4cm] {Info. proc. 9};

\fill[blue!20!white, xshift=7cm, yshift=-1.5cm] (0, 0) rectangle (1, 1);
\fill[green!20!white, xshift=7cm, yshift=-1.5cm] (0, 1) rectangle (1, 2);
\fill[blue!20!white, xshift=7cm, yshift=-1.5cm] (0, 2) rectangle (1, 3);
\fill[green!20!white, xshift=7cm, yshift=-1.5cm] (1, 0) rectangle (2, 1);
\fill[blue!20!white, xshift=7cm, yshift=-1.5cm] (1, 1) rectangle (2, 2);
\fill[green!20!white, xshift=7cm, yshift=-1.5cm] (1, 2) rectangle (2, 3);
\fill[blue!20!white, xshift=7cm, yshift=-1.5cm] (2, 0) rectangle (3, 1);
\fill[green!20!white, xshift=7cm, yshift=-1.5cm] (2, 1) rectangle (3, 2);
\fill[blue!20!white, xshift=7cm, yshift=-1.5cm] (2, 2) rectangle (3, 3);

\draw[step=1cm, yshift=.5cm] (7,-2) grid (10, 1);

\draw[-to] (feature.east) -- (filter1.west);
\draw[-to] (feature.east) -- (filter2.west);
\draw[-to] (feature.east) -- (filter3.west);
\draw[-to] (feature.east) -- (filter4.west);
\draw[-to] (feature.east) -- (filter5.west);
\draw[-to] (feature.east) -- (filter6.west);
\draw[-to] (feature.east) -- (filter7.west);
\draw[-to] (feature.east) -- (filter8.west);
\draw[-to] (feature.east) -- (filter9.west);

\draw[-to] (filter1.east) -| (7.5, 1.2);
\draw[-to] (filter2.east) -| (8.5, 1.2);
\draw[-to] (filter3.east) -- (9.5, 1.2);
\draw[-to] (filter4.east) -| (7.5, 0.2);
\draw[-to] (filter5.east) -- (8.4, 0);
\draw[-to] (filter6.east) -| (9.5, -0.2);
\draw[-to] (filter7.east) -- (7.5, -1.2);
\draw[-to] (filter8.east) -| (8.5, -1.2);
\draw[-to] (filter9.east) -| (9.5, -1.2);

\node at (7.5, 1) {1};
\node at (8.5, 1) {2};
\node at (9.5, 1) {3};
\node at (7.5, 0) {4};
\node at (8.5, 0) {5};
\node at (9.5, 0) {6};
\node at (7.5, -1) {7};
\node at (8.5, -1) {8};
\node at (9.5, -1) {9};
\end{tikzpicture}}
    \caption{Information processing in an expansion layer.}
    \label{fig:expansion}
\end{figure}

As shown in \autoref{fig:risnet_partial},
in a \gls{ris} of shape $36 \times 36 $
(1296 elements),
we have 16 \gls{ris} elements (the blue dots) 
with the ability to estimate the channel gain from pilot signals sent by users.
Channel gains of the other \gls{ris} elements are unknown.
The proposed hardware is similar to hybrid \gls{ris}~\cite{ju2024beamforming},
in which a few active elements with RF chains can amplify signals,
but is simpler for implementation
because the elements only estimate the channel but does not amplifying the signal.
The information processing of a normal layer
(see \autoref{fig:risnet_partial})
is described by \eqref{eq:layer_processing_noma}.
In an expansion layer,
the considered \gls{ris} elements are extended from~1 to~9.
As illustrated in \autoref{fig:expansion},
for class~\texttt{c} and class~\texttt{a},
we define 9 information processing units
instead of 1 in a normal layer.
The outputs of them are the features of the adjacent \gls{ris} elements
for the next layer.
For example,
the output of information process unit~1 is the feature of the
\gls{ris} element to the upper-left corner of the original \gls{ris} element,
whereas
the output of information process unit~5 is the feature of the
original \gls{ris} element itself.
With two expansion layers,
the considered \gls{ris} elements are increased from
the 9~\gls{ris} elements with \gls{csi}
to all the \gls{ris} elements.

Formally, the output of \gls{ris} element~$n$ using information processing unit~$j$ in \gls{noma} is
\begin{equation}
\mathbf{f}_{\nu(n, j), i + 1} =
\begin{pmatrix}
     \text{ReLU}(\mathbf{W}^{\texttt{c}}_{i, j} \mathbf{f}^{\text{NOMA}}_{n, i} + \mathbf{b}_{i, j}^{\texttt{c}}) \\
     \left(\sum_{n'}\text{ReLU}(\mathbf{W}^{\texttt{a}}_{i, j} \mathbf{f}^{\text{NOMA}}_{n', i} + \mathbf{b}_{i, j}^{\texttt{a}})\right) \big/ N\\
\end{pmatrix},
\label{eq:expansion_layer_processing_noma}
\end{equation}
where $\nu(n, j)$ is the \gls{ris} element index when applying information processing unit~$j$ for input of \gls{ris} element~$n$.
According to \autoref{fig:risnet_partial} and assuming that the \gls{ris} element index begins with~1 at the upper left corner,
increases first along the row and then changes to the next row 
(i.e., the index in row~$w$ and column~$h$ is ${h + (w - 1)\cdot H}$, with $H$ being the number of columns of the \gls{ris} array),
we have
\begin{equation}
    \nu(n, j)= 
\begin{cases}
n - H - 2 + j & j = 1, 2, 3,\\
n - 5 + j & j = 4, 5, 6,\\
n + H - 8 + j & j = 7, 8, 9.\\
\end{cases}
\label{eq:nu}
\end{equation}
\subsection{Hybrid Solution of Analytical Precoding and ML-enabled RIS Configuration}

Problem~\eqref{eq:problem_noma} involves the joint optimization of $\mathbf{V}$ (precoding)
and $\boldsymbol{\Phi}$ (\gls{ris} configuration).
While
the \gls{ris} configuration is optimized by the proposed RISnet,
the precoding problem has been intensively studied with known optimal solution~\cite{chen2016optimal,Chen22016}.


Exploiting the advantage of RISnet and the 
optimality of the precoding in quasi-degraded channels,
we use RISnet to configure the \gls{ris} and
use the optimal precoding in quasi-degraded channels for the \gls{bs}~\cite{Chen22016}.
The objective is to minimize the transmit power~\eqref{eq:optProbMinPW} subject to the \gls{qd} constraint,
defined as
\begin{equation}
    \norm{\mathbf{v}_1} + \norm{\mathbf{v}_2} + \eta p,
    \label{eq:training_objective}
\end{equation}
where $\eta$ is a large constant to ensure that the channel is quasi-degraded
and $p$ is the penalty if the channel is not quasi-degraded.
The definition of $p$ can be found in~\cite{Chen22016}.
Note that the training objective~\eqref{eq:training_objective} is differentiable.
Therefore,
the precoding is treated as part of the differentiable objective function,
allowing for gradient-based \gls{nn} training.

To summarize the proposed approach,
the training process is formulated in \autoref{alg:noma}.

\begin{algorithm}
\caption{RISnet training}
\label{alg:noma}
\begin{algorithmic}[1]
\State Initialize the RISnet $N_\theta$.
\Repeat
\State Randomly choose data samples in a batch.
\State Compute phase shifts $\boldsymbol{\Phi}$ with current $N_\theta$.
\State Compute the \gls{qd} penalty for every data sample.
\State Compute precoding vectors for every data sample,
where the precoding vectors are considered as functions of $\theta$.
\State Compute objective \eqref{eq:training_objective} with \gls{csi}, rate requirements, precoding (parameterized by $\theta$) and phase shifts.
\State Compute gradient of \eqref{eq:training_objective} \gls{wrt} $\theta$.
\State Perform an optimization with the gradient.
\Until{convergence}
\end{algorithmic}
\end{algorithm}

\section{Training and Testing Results}
\label{sec:results}

\begin{figure}
    \centering
    \resizebox{.35\linewidth}{!}{		\begin{tikzpicture}[scale=1]
            \tikzstyle{base}=[isosceles triangle, draw, rotate=90, fill=gray!60, minimum size =0.12cm]
   
			\foreach \i in {1, 1.8, 3.4, 4.2, 5.0, 5.8}
			\foreach \j in {2, 4}
			{
				\fill[gray!30!white] (\i, \j) rectangle (\i+0.6, \j+1);
			}
   
			\foreach \i in {1.4, 3.4}
			\foreach \j in {-0.4, 0.8}
			{
				\fill[gray!30!white] (\i, \j) rectangle (\i+1, \j+1);
			}
			\node[align=center, rectangle,draw, minimum height=.65cm, text width=1cm] (ue) at (4.7,3.5) {Users};
			\draw (1,3.9) -- (2.5,3.9);
			\draw (1,3.1) -- (2.5,3.1);
			\draw (3.3,3.9) -- (6.4,3.9);
			\draw (3.3,3.1) -- (6.4,3.1);
			
			\draw (2.5,-0.4) -- (2.5,3.1);
			\draw (3.3,-0.4) -- (3.3,3.1);
			\draw (2.5,3.9) -- (2.5,5);
			\draw (3.3,3.9) -- (3.3,5);
			
            \node[base] (BS) at (2.45,0.1){};
			\node[below of=BS, yshift=.2cm] (bs) {BS};
            \draw[decoration=expanding waves,decorate] (BS) -- (3.2,1.9, 1.2);
			\node[rectangle, fill=white, draw, scale=.8] (ris) at (3.6,4.1)
			{  \hspace{.0cm}RIS\hspace{.0cm} };
	\end{tikzpicture}}
    \vspace*{-.5em}
    \caption{The considered scenario: an intersection in an urban environment.}
    \label{fig:scenario}
\end{figure}
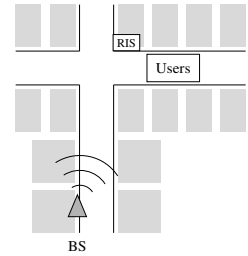

We use the open-source DeepMIMO data set~\cite{Alkhateeb2019}
to create training and testing sets.
In the scenario shown in \autoref{fig:scenario},
the \gls{los} channel from \gls{bs} to users is blocked by a building.
Only a weak direct channel is available through reflections on buildings and ground,
such that the \gls{ris} plays a crucial role in the transmission.
The user grouping is assumed to be given.
Important assumptions and parameter settings are listed in \autoref{tab:params}.

\begin{table}[htbp]
    \centering
    \caption{Scenario and model parameters}
    \label{tab:params}
    \begin{tabularx}{\linewidth}{lX}
        \toprule
        Parameter & Value \\
        \midrule
        Number of \gls{bs} antennas & \num{9} \\
        \gls{ris} size & $36 \times 36$ elements\\
        Carrier frequency & $\SI{3.5}{\GHz}$\\
        Distance between adjacent antennas at \gls{bs} & 0.5 wavelength\\
        Distance between adjacent antennas at \gls{ris} & 0.25 wavelength\\
        Learning rate & $8 \times 10^{-4}$ -- $1.5\times 10^{-3}$\\
        Batch size & \num{512}\\
        Optimizer & ADAM\\
        Number of data samples in training set & \num{10240}\\
        Number of data samples in testing set & \num{1024}\\
        \bottomrule
    \end{tabularx}
\end{table}

As explained earlier,
the performance with partial \gls{csi}
strongly depends on the spatial correlation of the channel model.
We use three channel models to assess the feasibility of applying partial \gls{csi} for \gls{ris} configuration:
\begin{itemize}
    \item Deterministic ray-tracing channel from the DeepMIMO simulator
    with a strong spatial correlation.
    \item Deterministic ray-tracing channel plus \gls{iid} scattering gains on each \gls{ris} element.
    The spatial correlation is weaker due to the \gls{iid} scattering gains.
    \item \Gls{iid} channel model due to scattering of infinitely many infinitely weak propagation paths without correlation.
\end{itemize}
In the following,
we present training and testing results with these three channel models.

\subsection{Training Behavior}

\begin{figure}
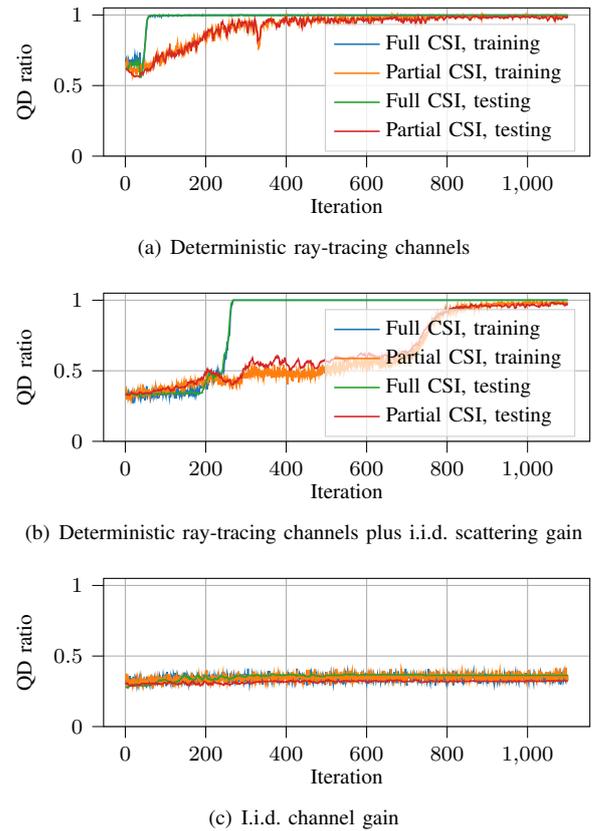

    \centering
    \subfigure[Deterministic ray-tracing channels]
    {\input{figs/noma_0_mutual_coupling_ratio}}
    \subfigure[Deterministic ray-tracing channels plus i.i.d. scattering gain]
    {\input{figs/noma_p_mutual_coupling_ratio}}
    \subfigure[I.i.d. channel gain]
    {\input{figs/noma_iid_mutual_coupling_ratio}\label{fig:ratio_noma_mutual_coupling_c}}
    \caption{Realized QD ratio in NOMA with mutual coupling in training and testing.}
    \label{fig:ratio_noma_mutual_coupling}
\end{figure}

\begin{figure}
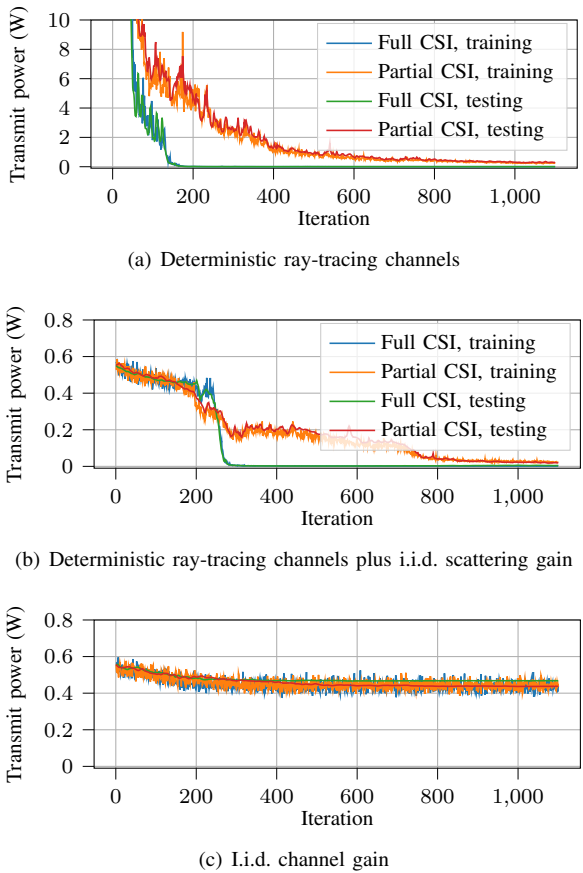

    \centering
    \subfigure[Deterministic ray-tracing channels\label{fig:det_power}]
    {\input{figs/noma_0_mutual_coupling_power}}
    \subfigure[Deterministic ray-tracing channels plus i.i.d. scattering gain\label{fig:semi_power}]
    {\input{figs/noma_p_mutual_coupling_power}}
    \subfigure[I.i.d. channel gain\label{fig:iid_power}]
    {\input{figs/noma_iid_mutual_coupling_power}\label{fig:power_noma_mutual_coupling_c}}
    \caption{Realized transmit power in NOMA with mutual coupling in training and testing. Please note the different scalings of the y axis.}
    \label{fig:power_noma_mutual_coupling}
\end{figure}

\autoref{fig:ratio_noma_mutual_coupling} and \autoref{fig:power_noma_mutual_coupling} show the improvement of \gls{qd} ratio and transmit power.
With the first two channel models,
the \gls{qd} ratios are almost 100\,\% at the end of training for both full \gls{csi} and partial \gls{csi}.
The training also reduces the transmit power significantly.
However,
the realized transmit power with full \gls{csi} is lower than the transmit power with
partial \gls{csi},
suggesting that having full \gls{csi} has its advantage in energy efficiency.
With \gls{iid} channel gains,
neither \gls{qd} ratio nor transmit power improve significantly during training
because the partial \gls{csi} does not provide sufficient information for the optimization.
Therefore, the optimization is only done to improve the average performance when the users are uniformly distributed in the given area.

According to~\cite{he2021wireless},
the wireless channel is \emph{sparse} in many typical scenarios,
i.e., the signal arrives at the receiver via a few specular \glspl{mpc}
and the spatially \gls{iid} scattering effect has very limited impact on the total channel gain.
This fact is the foundation of many compressed sensing based channel estimation algorithms, e.g.,~\cite{haghighatshoar2017massive,wunder2019low}.
Since most real wireless channels are similar to the first two channel models,
the proposed method with partial \gls{csi} is expected to work well not only in simulation,
but also in reality.

It is evident that the result with the most significance is
the one with deterministic channel model plus \gls{iid} scattering gain
because it is the realistic and more difficult scenario.
As described above,
the fully deterministic channel model is less challenging
while the \gls{iid} channel model is unrealistic and not suitable to apply partial \gls{csi}.
In the following,
we only present testing results with deterministic channel model plus \gls{iid} scattering gain (the second channel model).

\subsection{Comparison with Baselines and Necessity to Consider Mutual Coupling}

\begin{figure}
    \centering
    \begin{tikzpicture}
\definecolor{darkgray176}{RGB}{176,176,176}
\definecolor{steelblue31119180}{RGB}{31,119,180}
\tikzstyle{every node}=[font=\footnotesize]

\begin{axis}[
height=.5\linewidth,
width=.5\linewidth,
xbar,
x grid style={darkgray176},
xlabel={QD ratio},
xlabel near ticks,
y grid style={darkgray176},
symbolic y coords={
  {SDR (works only with 64 elements)~\cite{fu2019intelligent}},
  Random,
  {Proposed, partial CSI, w/o MC},
  {Proposed, full CSI, w/o MC},
  {Proposed, partial CSI},
  {Proposed, full CSI}
},
ytick=data,
xmin=0, xmax=1,
enlarge x limits={.1,upper},
nodes near coords,
axis y line*=none,
]
\addplot[xbar, fill=steelblue31119180] coordinates {
  (.95,{SDR (works only with 64 elements)~\cite{fu2019intelligent}})
  (.65,Random)
  (.96,{Proposed, partial CSI, w/o MC})
  (.93,{Proposed, full CSI, w/o MC})
  (.9961,{Proposed, partial CSI})
  (1.,{Proposed, full CSI})
};

\end{axis}
\end{tikzpicture}
    \vspace*{-1em}
    \caption{Test result of quasi-degradation ratio of models without mutual coupling and baselines.
    The SDR algorithm fails to optimize the RIS with more than 64 elements.}
    \label{fig:test_qd}
\end{figure}
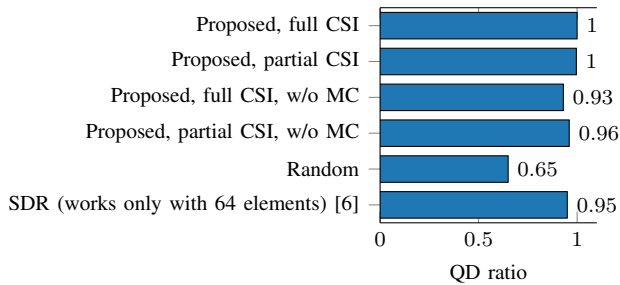

In this section,
we compare our proposed approach with baselines,
and results with and without consideration of mutual coupling.
Note that the test results are obtained with the testing set
sharing no common data samples with the training set,
i.e., the optimizer has not seen the testing data during training.
Since \gls{ris} optimization considering mutual coupling without partial \gls{csi}
for \gls{noma} is still an open problem,
we assume a \gls{ris} without mutual coupling with full \gls{csi}, and use
\gls{sdr}~\cite{fu2019intelligent} and
random phase shift
as baselines for comparison.
\autoref{fig:test_qd} shows the comparison of the \gls{qd} ratio with proposed approach and baselines.
From the first two rows,
we confirm that both RISnets with full \gls{csi} and partial \gls{csi} 
make all channels in the data set quasi-degraded.
Observing the third and fourth rows,
we realize that if we train the RISnets without considering mutual coupling 
and test them with mutual coupling,
they fail to realize a \gls{qd} ratio of 1.
This performance degradation due to model mismatch indicates the
necessity to consider mutual coupling if it exists.
The fifth and sixth rows of \autoref{fig:test_qd} show the realized
\gls{qd} ratios with random phase shifts of \gls{ris}
and using the \gls{sdr} algorithm~\cite{fu2019intelligent} as baselines.
Both baselines realize a lower \gls{qd} ratio than RISnet,
indicating a better performance of the proposed algorithm than state-of-the-art.
Besides the performance,
computing the \gls{ris} phase shifts with a trained RISnet takes a few milliseconds compared to a few seconds of the \gls{sdr} algorithm,
suggesting a high feasibility of the RISnet for real time application.
Note that the \gls{sdr} algorithm fails to configure a \gls{ris} with more than 64 elements,
confirming good scalability of our proposed approach.

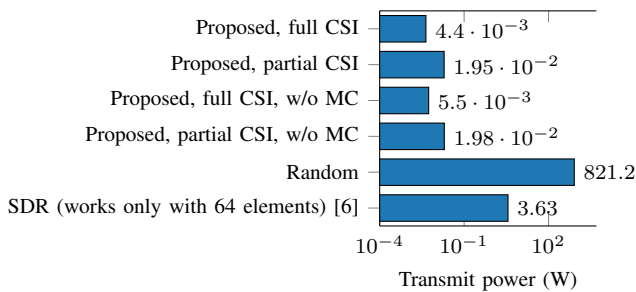
\begin{figure}
    \centering
    \begin{tikzpicture}

\definecolor{darkgray176}{RGB}{176,176,176}
\definecolor{steelblue31119180}{RGB}{31,119,180}
\tikzstyle{every node}=[font=\footnotesize]

\begin{axis}[
height=.5\linewidth,
width=.5\linewidth,
xbar,
xmode=log,
log origin=infty,
x grid style={darkgray176},
xlabel={Transmit power (W)},
xlabel near ticks,
xmin=1e-4, xmax=1000,
y grid style={darkgray176},
symbolic y coords={
  {SDR (works only with 64 elements)~\cite{fu2019intelligent}},
  Random,
  {Proposed, partial CSI, w/o MC},
  {Proposed, full CSI, w/o MC},
  {Proposed, partial CSI},
  {Proposed, full CSI}
},
ytick=data,
nodes near coords,
nodes near coords align={horizontal},
point meta=rawx,
enlarge x limits={.1,upper},
axis y line*=none,
]
\addplot[xbar,fill=steelblue31119180] coordinates{%
  (3.625,{SDR (works only with 64 elements)~\cite{fu2019intelligent}})
  (821.2,Random)
  (0.0198,{Proposed, partial CSI, w/o MC})
  (.0055,{Proposed, full CSI, w/o MC})
  (.0195,{Proposed, partial CSI})
  (.0044,{Proposed, full CSI})
};
\end{axis}

\end{tikzpicture}
    \vspace*{-1em}
    \caption{Test result of transmit power of models without mutual coupling and baselines.
    The SDR algorithm fails to optimize the RIS with more than 64 elements.}
    \label{fig:test_power}
\end{figure}

\autoref{fig:test_power} shows the realized transmit power of the proposed RISnet and baselines.
The full \gls{csi} realizes a lower transmit power than the partial \gls{csi}.
Moreover,
they both achieve a significantly better results
than the two baselines.

\section{Conclusion}

We have proposed the problem-specific \gls{nn} architecture RISnet
for \gls{ris} configuration and \gls{noma}.
Using unsupervised \gls{ml}
and analytical optimal precoding in quasi-degraded broadcast channels,
we realize a high energy efficiency given data rate requirements.
A significant advantage of the RISnet is its high scalability to control more than 1000 \gls{ris} elements
because the number of trainable parameters is independent from the number of \gls{ris} elements.
Moreover, the proposed RISnet requires partial \gls{csi} of only 16 \gls{ris} elements,
making the channel estimation much more feasible compared to full \gls{csi} requirement.
Furthermore, the short distance between \gls{ris} elements may result in mutual coupling between them,
which is addressed in our optimization.
Testing results show that the RISnet can realize quasi-degraded channels and low transmit powers,
outperforming state-of-the-art significantly
in both performance and computation time.

Code and data in this paper are publicly available under \url{https://github.com/bilepeng/risnet_noma_partial_csi}.

\printbibliography[heading=bibintoc]

\end{document}